\documentclass{article}

\usepackage[margin=1.4in]{geometry}
\usepackage{verbatim}
\usepackage{natbib}
\usepackage{amsfonts}
\usepackage{amssymb}
\usepackage{amsmath}
\usepackage{amsthm}
\usepackage{mathrsfs}
\usepackage[pdftex]{color,graphicx}
\usepackage{enumerate}
\usepackage{hyperref}
\usepackage{todonotes}
\usepackage{lscape}
\usepackage{comment}
\usepackage{setspace}

\DeclareMathOperator*{\argmax}{argmax\;}

\newtheorem{theo}{Theorem}

\newtheorem{lem}{Lemma}

\theoremstyle{remark}
\newtheorem{ex}{Example}
\newcounter{proofeq}

\def\bal#1\eal{\begin{align}#1\end{align}}
\def\bals#1\eals{\begin{align*}#1\end{align*}}
\def\be#1\ee{\begin{equation}#1\end{equation}}

\newcommand{\prog}{program}
\newcommand{\dm}{decision-maker} 
\newcommand{\opt}{option} 
\newcommand{\opts}{options} 
\newcommand{\feasActs}{\mathcal{A}} 
\newcommand{\out}{Y} 
\newcommand{\R}{R} 
\newcommand{\Act}{A} 
\newcommand{\act}{a} 
\newcommand{\oAct}{A^*} 
\newcommand{\eoAct}{p} 

\newcommand{\para}{\Theta} 
\newcommand{\epara}{\bar{\Theta}} 
\newcommand{\eparaopt}{\bar{\Theta}^*} 

\newcommand{\option}{j} 
\newcommand{\period}{t} 
\newcommand{\periods}{T} 
\newcommand{\batch}{M} 
\newcommand{\options}{J} 

\newcommand{\Exp}{\mathbf{E}} 
\newcommand{\Prob}{\mathbf{P}} 
\newcommand{\Filt}{\mathcal{F}} 
\newcommand{\KL}{D_{KL}} 
\newcommand{\Inf}{I} 
\newcommand{\Ent}{H} 

\newcommand{\irefs}{u} 
\newcommand{\ilocs}{v} 

\title{Adaptive Combinatorial Allocation
}

\author{Maximilian Kasy\footnote{Department of Economics,
University of Oxford, \texttt{maximilian.kasy@economics.ox.ac.uk}.} \and
Alexander Teytelboym\footnote{Department of Economics,
University of Oxford, \texttt{alexander.teytelboym@economics.ox.ac.uk}.}}

\begin{document}

\maketitle

\begin{abstract}
We consider settings where an allocation has to be chosen repeatedly, returns are unknown but can be learned, and decisions are subject to constraints.
Our model covers two-sided and one-sided matching, even with complex constraints. 
We propose an approach based on Thompson sampling.
Our main result is a prior-independent finite-sample bound on the expected regret for this algorithm. Although the number of allocations grows exponentially in the number of participants, the bound does not depend on this number. 
We illustrate the performance of our algorithm using data on refugee resettlement in the United States.
\end{abstract}


\onehalfspacing
\section{Introduction}
\label{sec:introduction}
Adaptive experimentation uses information obtained in the course of an experiment in order to optimize the treatment assignment for later study participants.\footnote{We thank Daniel Privitera and Manos Perdikakis for excellent research assistance.} For example, if job seekers arrive at a job center over time, a policymaker can use the outcomes of earlier job seekers in order to improve the assignment of labor market interventions for later participants \citep{caria2020adaptive}. 
Using this information, adaptive experimentation can for instance be used to maximize the welfare of study participants, or to inform subsequent policy choices \citep{kasy2019adaptive}.

In many policy settings, however, policymakers do not simply choose between a few interventions. Instead, they need to select an \emph{entire allocation of resources} among participants. These resources are typically scarce, and feasible allocations can be described by combinatorial constraints. 
For example, if the policymaker 
wants to allocate students to classrooms when classroom composition affects student outcomes
\citep{graham2010measuring}, she must ensure that all students are assigned to classrooms, that the capacity of classrooms is not exceeded, and that the allocation respect the demographic composition of students in the population. 
If the policymaker wants to 
allocate children to foster families when families impact the outcomes of the children, she needs to ensure that siblings are placed together and that foster homes are close to schools and family homes \citep{macdonald2019foster,robinson2019gets}. 
If the policymaker wants to allocate tenants to social housing, she needs to ensure that housing matches the needs of tenants and respects waiting-list priorities
\citep{thakral2016public,waldinger2018targeting,van2019socio}. 
If the policymaker wants to allocate combinations of therapies to different patients in order to overcome a disease, she needs to ensure that the therapies are actually available at the appropriate time and can be combined.

Combinatorial resource constraints might make
adaptive experimentation more difficult relative to the unconstrained case, as considered in the multi-armed bandit literature, since the number of possible allocations can be vast. For example, the number of ways to allocate students to classrooms grows exponentially in the number of students. This might cause both computational difficulties (requiring optimization over a large discrete space), and statistical difficulties (the expected rewards for many different allocations have to be learned).
We show, however, that remarkably despite these difficulties, performance close to the unconstrained case can be achieved.
In this paper we consider an adaptive allocation policy extending the idea of Thompson sampling \citep{thompson1933likelihood}.
This policy is close to optimal for maximizing the outcomes of experimental participants in the presence of
combinatorial resource constraints.

The general setting we consider is the following. The decision-maker has
access to a finite number of \emph{options} (e.g., matches) but is constrained to
selecting only allocations (combinations of options; e.g., matchings) that satisfy the resource constraints
(e.g., a one-to-one matching). Participants arrive in \emph{batches} every
period. The decision-maker selects an allocation and observes the
outcome of each selected option. The outcome of each option results in a \emph{reward}. The decision-maker's objective is to maximize the expected cumulative rewards from all the options she picked over time; equivalently, the decision-maker aims to minimize expected \emph{regret}, i.e., the expected difference between the optimal
combination of options in each period. This type of setting has been called a \emph{combinatorial
semi-bandit} setting with \emph{linear} rewards in the literature \citep{audibert2011minimax}. ``Combinatorial'' because the
decision-maker can choose combinations of options; ``semi-bandit'' because the
decision-maker can observe the outcomes of every match, not just of the entire
matching; and ``linear rewards'' because the objective function is the sum of
the rewards of all matches made.

Our main theoretical result is a bound on the worst-case regret obtained by using Thompson sampling in
our setting. 
Thompson sampling is a classic heuristic for standard bandit problems; it requires that each action is picked with probability equal to the posterior probability that this action is optimal. Our theoretical result is appealing for three reasons. First, the worst-case
expected regret does not depend on the batch size even though the number of possible
actions (i.e., allocation) grows exponentially in the batch size. Second, our bound holds in finite-samples and does not rely on asymptotic approximations. Third, our bound is prior-independent allowing for arbitrary statistical dependence across match outcomes; this is particularly relevant for matching contexts.\footnote{This contrasts with existing bounds which require prior independence across options, as discussed below.}

We apply our approach to the problem of matching resettled refugees to local communities in the United States \citep{bansak2018improving,trapp2018placement}. Our data cover the placement of all refugees by HIAS, an American resettlement agency, between 2014 and 2020. Our objective is to maximize the probability of employment of refugees in the first three months after their arrival. 
The allocation of refugees to local communities is subject to constraints.
Local communities have a quota on the total number of refugees they can resettle in a given year, and refugees who come from the same family must be placed together. The implied optimization problem is equivalent to a ``multiple knapsack problem'' \citep{trapp2018placement,delacretaz2019matching}.


%
%

Our paper is closely related to the literature on
multi-armed bandit problems. Rather than attempting to characterize analytical
solutions (e.g., \citet{gittins1979bandit}), we focus on analyzing properties of
the well-known probability matching heuristic due to
\citet{thompson1933likelihood}. 
\citet{agrawal2012analysis} and
\citet{kaufmann2012thompson} have shown, for the fixed parameter case, that the asymptotic bound on expected regret of the
Thompson algorithm in bandit settings matches the lower bound on regret for \emph{any} bandit
algorithm, which was derived by \citep{lai1985asymptotically}. \citet{wang2018thompson} provide a distribution-dependent regret bound for the Thompson algorithm in the \emph{combinatorial semi-bandit} setting; our result is distribution-free. Adaptive experimentation using the Thompson algorithm has been proposed for applications such as drug trials \citep{berry2006bayesian},
recommender systems \citep{kawale2015efficient}, and customer acquistion
\citep{schwartz2017customer}. More recently, adaptive experimentation has been
deployed in field experiments in development contexts
\citep{kasy2019adaptive,caria2020adaptive}.

The rest of the paper is organized as follows. Section~\ref{sec:setup} describes
our combinatorial semi-bandit setting and the Thompson heuristic; Section~\ref{ssec:examples} then discusses several examples covered by this general framework.
Section~\ref{sec:guarantees} gives our main theoretical result and the intuition for its
proof. Section~\ref{sec:implementation} covers several considerations for implementation in practice, including the choice of model and prior, methods for sampling from the posterior, and statistical inference. Section~\ref{sec:applications} discusses calibrated simulations based on the motivating applications. 
Appendix \ref{sec:information} provides a brief review of information theory, which is need for the proof of our main result.
All proofs can be found in Appendix~\ref{sec:proofs}.

\section{Setup}\label{sec:setup}

We denote all random variables with capital letters (e.g., $\Act$) and the realizations of random variables with lower-case letters (e.g., $\act$).

\paragraph{Feasible \opts\ and actions}
The \dm\ has access to \emph{\opts}\ (e.g., matches or projects) $\option \in 1,\ldots,\options$, but only has sufficient resources to select $\batch \leq \options$ of these.
We denote by $\feasActs \subseteq \{\act\in \{0,1\}^\options:\, \|\act\|_1 = \batch\}$ a collection of \emph{feasible} combinations of \opts.
$\feasActs$ is a strict subset if the \dm\ faces additional allocation constraints.
The \dm's \emph{action}
$\act \in \feasActs$ is a feasible combination of \opts.

\paragraph{Timing, potential outcomes, and observability}
The \prog\ takes places in a finite number of periods $\period = 1,\ldots, \periods$.
In each period, there is a vector $\out_\period \in [0,1]^\options$ of potential outcomes, where $\out_{\option \period}$ is the potential outcome for option $\option$ in period $\period$.
The vectors $\out_\period$ are i.i.d. across periods.
We denote the average potential outcome (or average structural function) for option $j$ by $\para_\option$, that is, $\para_\option=\Exp[\out_{\option\period} | \para]$.
The \dm\ holds a prior belief over the vector $\para \in [0,1]^\options$, where we allow for arbitrary dependence of this prior across the \opt s $\option$.

In each period, the \dm\ chooses an action $\Act_\period \in \{0,1\}^\options$.
If the \dm\ chooses action $\act$, they observe the outcomes of the chosen options $\option$ (the \opt s $\option$ for which $\act_\option =1$), i.e., the vector
\be
  \out_\period(\act) =(\act_{\option}\cdot \out_{\option \period} :\; \option = 1,\ldots,\options).
\ee

%

We assume ``stable unit treatment values'' (i.e., no spillovers or interference) across options $\option$, in the sense that $Y_{\option \period}$ does not depend on the chosen action $\act_{\option' \period}$ for any $\option'$.
Note that this assumption is consistent with settings where $\out_{\option \period}$ is itself the equilibrium outcome of interactions across multiple individuals comprising an \opt\ $\option$, 
as is the case for example in the applications to peer effects or matching discussed below.

Given our assumption about observability, the \emph{information} available at the beginning of period $\period$ is given by
\be
  \Filt_\period = \left\{
    \left ( \Act_{\period'}, \out_{\period'}(\Act_{\period'}) \right ) : \;
    1\leq \period' < \period
  \right\}.
\ee
Throughout this paper, the subscript $\period$ on $\Exp_\period$ indicates that the expectation is evaluated under the posterior distribution $\Prob_\period(\cdot) = \Prob(\cdot\mid\Filt_\period)$, where $\Filt_\period$ is the information available at the beginning of period $\period$.
The \dm\ can choose their action $\Act_\period$ at the beginning of each period $\period$ based on the information $\Filt_\period$, as well as possibly based on a randomization device that is statistically independent across periods and independent of the sequence of potential outcomes $\left ( \out_\period \right )_{\period = 1}^\periods $.

\paragraph{Objective and policy}

If the \dm\ chooses action $\act$ in period $\period$, they receive a \emph{reward} which is equal to $\langle \act, \out_\period \rangle$.
Therefore, upon taking action $\act$ the \dm's \emph{expected reward} given $\para$, which is the same across periods $\period$, equals
\be
 \R(\act) =\Exp_\period[\langle \act, \out_\period \rangle | \para]= \langle \act, \para \rangle.
\ee
The \dm ~would like to maximize cumulative expected rewards,
\be
  \Exp_1 \left [ \sum_{\period = 1}^\periods \R(\Act_\period) \right ].
\ee
The expectation in this expression is taken over the randomness in the choice of actions $\Act_\period$, the sampling distribution of potential outcomes $\out_t$, and over the prior distribution of $\para$.
Denote by $\oAct$ a feasible action that maximizes the expected reward conditional on $\para$ (but not conditional on the vector $\out_\period$), that is,
\be
  \oAct \in \argmax_{\act \in \feasActs} \R(\act) = \argmax_{\act \in \feasActs} \langle \act, \para \rangle.
\ee

The \dm's objective is equivalent to minimizing expected \emph{regret} at $T$
\be
  \Exp_1 \left [ \sum_{\period = 1}^\periods \left ( \R(\oAct) - \R(\Act_\period) \right )  \right ].
\ee

Solving this dynamic stochastic combinatorial optimization problem is computationally quite costly.
Rather than solving it, we propose that the \dm\ adopt the following heuristic policy.
In each period the \dm\ should take an action $\act$ from the feasible set $\feasActs$ according to the posterior probability that $\act$ is optimal, that is, for each $\act\in\feasActs$,
\be
  \Prob_\period(\Act_\period=\act)=
  \Prob_\period(\oAct_\period=\act).
  \label{eq:thompson}
\ee
This assumption implies in particular that
$
  \Exp_\period[\Act_\period] = \Exp_\period[\oAct].
$
This heuristic approach is known as Thompson sampling, and was originally introduced by \cite{thompson1933likelihood} for treatment assignment in adaptive experiments.

\subsection{Examples}
\label{ssec:examples}

In the following we discuss several examples that are covered by our general framework, and thus in particular by the regret bound provided in Theorem \ref{theo:bounding_thompson} below.
These examples correspond to practically relevant policy problems. They also illustrate how various combinatorial allocation problems that have been studied in the literature fit into our framework, such as assignment to peers, one-to-one matching, many-to-one matching, knapsack problems, etc.

For each of these examples, several options might correspond to same underlying parameter, so that $\para_j = \para_{j'}$ with prior probability $1$, for some $j, j'$. In the case of one-to-one matching, for instance, each matched pair corresponds to one option, but $\para_j$ is the same for all matched pairs $j$ with the same observed covariates on both sides of the match.

\begin{ex}[Allocation of refugees to local communities]
American refugee resettlement agencies need to make weekly decisions about the allocation of arriving refugee families to local communities. 
An action $a$ is a matching of refugee families to local communities. The number of options $\options$ is the number of distinct matches between different family-locality pairs, and the batch size $\batch$ is equal to the number of refugee families arriving in a given week.
We will consider this example in greater detail in Section \ref{sec:applications} below.
\end{ex}

\begin{ex}[Foster parent allocation]
Foster families are typically able to host several foster children at the same time \citep{macdonald2019foster,robinson2019gets}. 
An action $\act$ is a \emph{many-to-one matching} between families and children. The feasible actions $\act$ require that no family receives more children than it can host, that all siblings are matched to the same foster family, and that children are hosted near their school and activities.
The parameters $\para_\option$ are again perfectly dependent across options $\option$ that are observationally identical, i.e., across matches of children and families with the same observed covariates.
\end{ex}

\begin{ex}[Peer effects and classroom composition]
Suppose that a policymaker would like to choose the gender composition of classrooms in order to maximize student performance \citep{graham2010measuring}. Assume students are of two types, boys and girls. Classrooms have a fixed number of students. 
An action $\act$ allocates (i.e., groups) the students into classrooms. Classroom identity does not matter, but the identity of peers does matter, for student outcomes.
The number of options $\options$ is equal to the number of classroom-sized subsets of the set of all students. The batch size $\batch$ is equal to the number of classrooms.
If students are observationally indistinguishable from each other, except for gender, then the prior exhibits perfect dependence across classrooms with the same number of girls and boys.
\end{ex}


%
%

\section{Performance guarantee}
\label{sec:guarantees}


We next state our main theoretical result.
This result provides tight worst-case guarantees for the expected regret of Thompson sampling in our setup.

\begin{theo}
  \label{theo:bounding_thompson}
  Under the assumptions of Section \ref{sec:setup},
  $$\Exp_1 \left [ \sum_{\period = 1}^\periods \left ( \R(\oAct) - \R(\Act_\period) \right )  \right ]\leq 
  \sqrt{\frac{1}{2}\options \periods  
  \batch \cdot \left [ \log\left ( \tfrac{\options}{\batch} \right ) +1 \right ] }.$$
\end{theo}


\paragraph{Discussion of Theorem \ref{theo:bounding_thompson}}
Several features of the regret  bound in Theorem \ref{theo:bounding_thompson} are worth emphasizing.
First, this bound is a finite sample bound. There is no large sample limit and no remainder term. 
Second, this bound does not depend on the prior distribution for $\para$ in any way. Furthermore, it allows for prior distributions with arbitrary statistical dependence across the components of $\para$, as required by our motivating examples.
Third, this bound implies that Thompson sampling in our setting achieves the efficient rate of convergence for regret: As shown by \cite{audibert2014regret}, the minimax regret in our setting grows at a rate of $\sqrt{\options \periods \batch}$, up to logarithmic terms.

Theorem \ref{theo:bounding_thompson} bounds the worst case expected regret across all possible priors, summed across units.
To get the worst case expected regret per unit, divide this expression by $\periods \batch$, which yields the bound 
$\sqrt{\options  \cdot \left [ \log\left ( \tfrac{\options}{\batch} \right ) +1 \right ] \big / \left ( 2 \periods \batch \right ) }$.
This bound goes to $0$ at a rate of $1$ over the square root of the sample size, that is, at a rate of $1/\sqrt{\periods \batch}$.
The theorem furthermore shows that this worst case expected regret grows, as a function of the number of possible options $\options$, like $\sqrt{\options}$ (neglecting the logarithmic term).
Remarkably, worst case regret does not grow in the batch size $\batch$.
This is despite the fact that the setup of Section \ref{sec:setup} allows for action sets of size ${\options}\choose{\batch}$.
For comparison, application of the worst case regret bound for Thompson sampling in bandits with dependent arms provided by Proposition 3 in \cite{russo2016information} yields a much larger bound which grows in proportion to $\sqrt{{\options \choose \batch}  \log  {\options \choose \batch}  }$.
Instead, the regret bound in Theorem \ref{theo:bounding_thompson} grows like that for a simple multiarmed bandit with $\options$ arms.

\paragraph{Intuition for the proof of Theorem \ref{theo:bounding_thompson}}

The proof of Theorem \ref{theo:bounding_thompson} is provided in Appendix \ref{sec:proofs}.
This proof builds on several definitions and standard results from information theory which are reviewed in Appendix \ref{sec:information}.
Here we just sketch some of the key steps in our proof.

First, we use Pinsker's inequality in order to relate expected regret to the information about the optimal action $\oAct$ provided by observations, where information is measured by the KL-distance of posteriors and priors.
Pinsker's inequality implies, for Bernoulli random variables $B$ and $B'$, that $(\Exp[B] - \Exp[B'])^2 \leq \tfrac{1}{2} \KL(B,B').$ 
Lemma \ref{lem:bounding_componentwise} applies Pinsker's inequality to terms showing up in the definition of expected regret which are of the form $\Exp_\period[\para_\option | \oAct_\option = 1] - \Exp_\period[\para_\option]$.
This use of Pinsker's inequality is at the core of the proofs in \cite{russo2016information}.

Second, following some of the ideas introduced in \cite{bubeck2020first}, Lemma \ref{lem:boundingKL} relates the KL-distance to the entropy of the events $\oAct_\option = 1$.
The combination of these two Lemmas allows to bound the expected regret for option $\option$ in terms of the entropy reduction for the posterior of $\oAct_\option$.

Third and lastly, Lemma \ref{lem:bounding_sum} shows that the total reduction of entropy across the options $\option$, and across the time periods $\period$, can be no more than the sum of the prior entropy for each of the  events $\oAct_\option = 1$, which is bounded by $\batch \cdot \left [ \log\left ( \tfrac{\options}{\batch} \right ) +1 \right ] $.
The proof of Theorem \ref{theo:bounding_thompson} then combines these three Lemmas.

\paragraph{Relationship to the literature}
Our proof builds on the information theoretic approach pioneered by \cite{russo2016information} (in particular Lemma 1 and 2 as well as Proposition 6 therein) and some variations of this approach proposed by \cite{bubeck2020first} (in particular Lemma 13).
Despite the close relationship to their arguments, Theorem \ref{theo:bounding_thompson} differs from the bounds provided in these papers as follows.
The closest result in \cite{russo2016information} is their Proposition 6. Their result, however, requires statistical independence of the prior and posterior distribution for the components of $\para$ at all times $\period$.
By contrast, Theorem \ref{theo:bounding_thompson} above allows for arbitrary dependence.
This is especially relevant for the matching setting, where independence in the prior distribution would be quite hard to justify.

The closest result in \cite{bubeck2020first} is their Theorem 21. Their result is asymptotic, rather than providing an exact finite sample bound. 
The main interest of \cite{bubeck2020first} is an asymptotic refinement of regret bounds that scales in the best achievable regret, allowing for the latter to converge to $0$; this is something which our result does not aim to do.

%
%
%


\section{Implementation of Thompson sampling for matching problems}
\label{sec:implementation}

\subsection{Model and prior for matching settings}
\label{ssec:model_prior}

In order to achieve good performance in practice, our proposed procedure relies on specifying an appropriate model for the data generating process, and an appropriate prior distribution for the underlying parameters.
We generally advocate for the use of default priors that are diffuse and symmetric across types, while incorporating reasonable assumptions about the dependency structure between different options $\option$.

Table \ref{tab:priors} proposes some variants of models and priors for matching settings, covering our leading motivating examples, including those used in our empirical applications.
For each of these variants, we assume that the options $\option$ consist of two-sided matches between types $\irefs_\option$ and types $\ilocs_\option$.
For each possible match (option), the potential outcomes $\out_{\option \period}$ are drawn from some distribution with mean $\para_{\option}$.
We need to specify this distribution of $\out_{\option \period}$, as well as a joint prior distribution of the parameters $\para_{\option}$ across $\option$.

Each of these models assumes that the option-effect $\para_{\option}$ is determined by the sum of type-effects ${\Gamma^\irefs}_{\irefs_\option}$ and ${\Gamma^{\ilocs}}_{\ilocs_\option}$, plus an interaction effect ${\Gamma^{\irefs \ilocs}}_{\irefs_\option, \ilocs_\option}$.
For continuous outcomes, we assume that $\para_{\option}$ is directly given by this sum.
For binary or discrete outcomes, we assume that $\para_{\option}$ is given by the logit link function applied to this sum.
 
For the model for outcomes with discrete bounded support, the distribution of $\out_{\option \period}$ is governed by the mean parameter $\para_{\option}$ as well as a dispersion parameter $m$. The latter is necessary to allow for larger dispersions relative to a more restrictive Binomial model, which might put excessive weight on the information content of single observations.
 
\begin{table}
\caption{Models and priors for matching}
\label{tab:priors}  
\rule{\textwidth}{0.4pt}  
\textbf{Continuous outcomes}
\bals
  \out_{\option \period}  &\sim N(\para_{\option}, \sigma^2) \\
  \para_{\option} &= {\Gamma^\irefs}_{\irefs_\option} + {\Gamma^{\ilocs}}_{\ilocs_\option} + {\Gamma^{\irefs \ilocs}}_{\irefs_\option, \ilocs_\option}\\
  {\Gamma^{\irefs}}_{\irefs_\option} &\sim N(0, \tau^2_{\Gamma^\irefs}),\quad
  {\Gamma^{\ilocs}}_{\ilocs_\option} \sim N(0, \tau^2_{\Gamma^\ilocs}), \quad
  {\Gamma^{\irefs \ilocs}}_{\irefs_\option, \ilocs_\option} \sim N(\mu, \tau^2_{\Gamma^{\irefs \ilocs}}),
\eals
\rule{\textwidth}{0.4pt}

\vspace{5pt}
\textbf{Binary outcomes}
\bals
  \out_{\option \period}  &\sim \textrm{Bernoulli}(\para_{\option}) \\
  \para_{\option} &= \frac{1}{1+\exp\left (- \left ( {\Gamma^{\irefs}}_{\irefs_\option} + {\Gamma^{\ilocs}}_{\ilocs_\option} + {\Gamma^{\irefs \ilocs}}_{\irefs_\option, \ilocs_\option} \right ) \right )}  \\
  {\Gamma^{\irefs}}_{\irefs_\option} &\sim N(0, \tau^2_{\Gamma^\irefs}),\quad
  {\Gamma^{\ilocs}}_{\ilocs_\option} \sim N(0, \tau^2_{\Gamma^\ilocs}), \quad
  {\Gamma^{\irefs \ilocs}}_{\irefs_\option, \ilocs_\option} \sim N(\mu, \tau^2_{\Gamma^{\irefs \ilocs}}),
\eals
\rule{\textwidth}{0.4pt}

\vspace{5pt}
\textbf{Discrete outcomes with bounded support $\{0,\ldots, \bar y\}$}
\bals
  \out_{\option \period}  &\sim \textrm{Beta-Binomial}(  \alpha_{\option}, \beta_{\option}, \bar y) \\
  A_{\option} &= m \cdot \para_\option, \quad
  B_{\option} = m \cdot (1-\para_\option) \\
  \para_{\option} &= \frac{1}{1+\exp\left (- \left ( {\Gamma^{\irefs}}_{\irefs_\option} + {\Gamma^{\ilocs}}_{\ilocs_\option} + {\Gamma^{\irefs \ilocs}}_{\irefs_\option, \ilocs_\option} \right ) \right )}  \\
  {\Gamma^{\irefs}}_{\irefs_\option} &\sim N(0, \tau^2_{\Gamma^\irefs}),\quad
  {\Gamma^{\ilocs}}_{\ilocs_\option} \sim N(0, \tau^2_{\Gamma^\ilocs}), \quad
  {\Gamma^{\irefs \ilocs}}_{\irefs_\option, \ilocs_\option} \sim N(\mu, \tau^2_{\Gamma^{\irefs \ilocs}}),
\eals
\rule{\textwidth}{0.4pt}
\footnotesize
\begin{flushleft}
\textit{Notes:}
For each of these cases we assume that the components of ${\Gamma^{\irefs}},{\Gamma^{\ilocs}},{\Gamma^{\irefs \ilocs}}$ are mutually independent given the hyper-parameters.
The hyper-parameters are given by  $\sigma^2,\tau^2_{\Gamma^\irefs},\tau^2_{\Gamma^\ilocs},\tau^2_{\Gamma^{\irefs \ilocs}}$ and $\mu$ for continuous outcomes,
by $\tau^2_{\Gamma^\irefs},\tau^2_{\Gamma^\ilocs},\tau^2_{\Gamma^{\irefs \ilocs}}$ and $\mu$
for binary outcomes, and by
$\tau^2_{\Gamma^\irefs},\tau^2_{\Gamma^\ilocs},\tau^2_{\Gamma^{\irefs \ilocs}}$ and $\mu$ for discrete outcomes with bounded support.
We propose to use some diffuse prior for these hyper-parameters.
\end{flushleft}
\end{table}

\subsection{Inference}

In order to implement Thompson sampling, we need to sample from the posterior for $\para$.
This posterior is also relevant for statistical inference on parameter values.
Such inference is often a secondary goal, in additional to the primary goal of maximizing participant outcomes.
Such inference might be Bayesian, using the same posterior distributions that go into the assignment algorithm.
Alternatively, such inference might be based on permutation tests, as described below.

\paragraph{Markov Chain Monte Carlo and Bayesian inference}
For hierarchical priors, such as those discussed in Section \ref{ssec:model_prior}, posterior distributions are not available in closed form, in general.
We can, however, sample from the posterior for $\para$ using Markov Chain Monte Carlo (MCMC) methods.
Such MCMC methods only require us to specify the posterior up to a multiplicative constant (typically, up to the denominator of the posterior density, which is given by the marginal density of the observed data).
MCMC methods are based on constructing a Markov Chain which converges to an ergodic distribution that is given by the posterior of interest.
There are various ways of constructing such Markov Chains; one of them is  Hamiltonian Monte Carlo.
In our applications, we sample from the posterior using Hamiltonian Monte Carlo as implemented in the software STAN \citep{carpenter2017stan}.

Let $\hat \para_\period$ be a draw from the posterior given $\Filt_\period$ generated by MCMC, after a sufficiently long warm-up period. Choose
\be
  \Act_\period = \argmax_{\act \in \feasActs} \langle \act, \hat \para_\period \rangle.
\ee
Then $\Act_\period$ follows the distribution required for Thompson sampling, that is, it satisfies Equation~\eqref{eq:thompson}.

In order to form $1-\alpha$ credible sets for the parameters $\para_\option$ given the history $\Filt_\period$, sample a large number of draws $\hat \para_\period$ from the posterior, and form a credible interval based on the $\alpha/2$ and $1-\alpha/2$ quantiles of $\hat \para_{\option, \period}$ across these draws.


\paragraph{Randomization inference}

An alternative to Bayesian inference is randomization (permutation) inference.
In the context of treatment effect estimation, randomization inference can be used to test the null hypothesis that treatment does not affect any outcome, so that for instance $Y_i^1 = Y_i^0$ for all units $i$ and treatment value $0,1$.

In the context of our setting, we need to modify this null hypothesis.
Permutation inference requires that we specify the counterfactual outcome vector $\out_\period^0(\act)$ for any counterfactual action $\act \in \feasActs$ under the null hypothesis $H^0$, given knowledge of $\out_\period(\Act_\period)$ for the realized action $\Act_\period$.
In many cases of interest, there might be more than one plausible way to specify such a null hypothesis and the corresponding counterfactual outcome vectors.

To illustrate, consider the case of one-to-one matching (of refugees to local communities, say), where each option $\option$ corresponds to a match of a refugee family to a local community.
We could formalize the null hypothesis that ``the matching does not matter'' in two different ways. We could consider the hypothesis that refugee outcomes are the same, no matter which community they are allocated to.
Or we could consider the hypothesis that outcomes in a community are the same, no matter which refugees are allocated to be there.

Given some specification of counterfactual outcomes, we can sample counterfactual histories $\tilde \Filt_\period$ by re-running the Thompson sampling algorithm iteratively. In each period $s$, draw $\tilde \para_{\period'}$ and the corresponding $\tilde \Act_{\period'}$ from the posterior given $\tilde \Filt_{\period'}$. Impute a counter-factual outcome vector $\out_{\period'}^0(\tilde \Act_{\period'})$, based on the null hypothesis to be tested. Update the history $\Filt_{\period'}$ by adding $\tilde \Act_{\period'}, \out_{\period'}^0(\tilde \Act_{\period'})$, and iterate for the next period.
Once $\period'=\period$, calculate a realization of the test-statistic as a function of $\tilde \Filt_\period$.
Repeat this process to generate a sampling distribution of the test-statistic, and corresponding critical values and p-values for testing the null hypothesis under consideration.

\section{Applications}
\label{sec:applications}

\subsection{Refugee resettlement}

The United States has historically been the world's largest destination of \textit{resettled} refugees, with 78,340 admitted in 2016. There is a lot of evidence that the initial match between refugees and local communities dramatically affects the socioeconomic outcomes of refugees \citep{bansak2018improving,trapp2018placement}. However, local community capacities are tightly regulated by the US government. As a result, one US resettlement agency (HIAS) optimizes the placement of the resettled refugees using its recommendation system called Annie MOORE. However, Annie's estimates of refugee employment are static and come from a LASSO regression run annually \citep{trapp2018placement}.

Here, we draw on the data used by Annie MOORE in order to run calibrated simulations for our proposed procedure subject to realistic constraints, hoping to inform actual refugee placement by Annie MOORE in the future.

\paragraph{Data}
Our data covers all refugees resettled by HIAS between October 2005 and September 2020. There were a total of 37,149 refugees that constitute 15,523 cases (i.e., families). We focus on the employment of primary applicants in each family. For each primary applicant in the family, we observe three binary variables: whether the applicant is of prime working age (25-54), their gender, and whether they are English-speaking. We also observe the family size, and the affiliate where the family was resettled. We furthermore observe whether the primary applicant had any US ties. Applicants with US ties (e.g., US resident friends or family) are automatically resettled to the community where their US ties reside. Applicants without US ties can be resettled to any of the communities where HIAS operates. Finally, we observe whether or not the primary applicant was employed within 90 days of arrival. This is a key metric used by the State Department to assess the performance of American resettlement agencies.

There are 57 affiliates in our data. We drop any affiliate with fewer than 150 resettled cases over the whole period under consideration, leaving us with 17 affiliates.\footnote{In their analysis, \citet{trapp2018placement} also pool some affiliates because of small numbers of observations.} All affiliates are anonymized.
Based on the available observables, we classify refugees into 8 ``types'' $u$, while treating each of the 17 affiliates as a separate ``type'' $v$. This means that there are $8 \cdot 17 = 136$ parameters (probabilities of finding employment) that we might wish to learn. 


As noted above, the affiliates have a limited capacity in hosting refugees, where capacity is the total number of resettled people (primary applicants and their family members).
The  capacities of affiliates can sometimes change throughout the course of the year.
For our simulations, we conservatively set the available annual capacities to be 110\% of the total number of refugees without US ties actually resettled to each affiliate in a given year.\footnote{Resettlement agencies are allowed to exceed their official capacity by 10\% without further approval; however, in practice, agencies occasionally have to seek approval for capacity extensions and reductions because of stochastic refugee flows. Dynamic capacity management is beyond the scope of this paper.} The monthly quota for each affiliate is therefore 110\% of its annual capacity divided by 12.

\paragraph{Simulation design}
Our simulated matching process works as follows.
For each month $\period$ in the available data, we consider all the refugees who were resettled by HIAS in this month. 
We match the refugees with US ties to their actual affiliates. 
For all the refugees without US ties, we match them to affiliates using the Thompson algorithm discussed above.
This matching has to satisfy the capacity constraints of affiliates described above, and the varying sizes of refugee families. Solving for the optimal matching for a given draw $\hat \para$ from the posterior defines a so-called multiple knapsack problem.
Any leftover capacity and any unmatched refugees without US ties are carried over to the next month.

We then impute counterfactual outcomes for the refugees allocated using the Thompson algorithm. based on calibrated parameter values $\para_0$. These calibrated parameter values are set to the posterior mean for the Bayesian hierarchical model for binary outcomes described in Section \ref{ssec:model_prior}.

\paragraph{Results}

WORK IN PROGRESS.

%


\clearpage
\bibliographystyle{apalike}
\bibliography{adaptive_matching}

\clearpage
\appendix
\renewcommand{\theequation}{\thesection\arabic{equation}}

\setcounter{equation}{0}
\section{A brief review of information theory}
\label{sec:information}

In this section we review some basic definitions and facts about entropy, mutual information, and KL-divergence.
For further background, see \cite{mackay2003information} (in particular chapter 8), as well as Section 3 in \cite{russo2016information}.
For our purposes, it is enough to restrict attention to the Bernoulli case, so that we can introduce the following definitions in elementary form.
Let $\Act$ be a Bernoulli random variable with expectation $\eoAct$, and let $\Act'$ be a Bernoulli random variable with expectation $q$.
We overload notation by allowing the arguments $\Act$ and $\eoAct$ to be used interchangeably.

\begin{itemize}
  \item \textbf{Entropy}:
  \be
    \Ent(\Act) = \Ent(\eoAct) = -\left [ \eoAct \log(\eoAct) + (1-\eoAct) \log (1-\eoAct) \right ]. 
  \ee
  \item \textbf{KL divergence}:
  \be
    \KL(\Act,\Act') = \KL(\eoAct,q) = \eoAct \log \left ( \tfrac{\eoAct}{q} \right ) + (1-\eoAct) \log \left ( \tfrac{1-\eoAct}{1-q} \right ) .
  \ee
  \item \textbf{Pinsker's inequality}:
  \be
  \left ( \Exp[\Act] - \Exp[\Act'] \right )^2  = |\eoAct-q| \leq  \tfrac{1}{2} \KL(\eoAct,q)= \tfrac{1}{2} \KL(\Act,\Act').
  \ee
  \item \textbf{Mutual information} as expected divergence of the posterior:\\
  For any random variable or vector $F$, let $\eoAct(f) = \Exp[\Act|F=f]$.
  Then
  \be
    \Inf(\Act;F)= \Exp[\KL(\eoAct(F),\eoAct)].
  \ee
  \item \textbf{Conditional entropy}:
  \be
  \Ent(\Act|F) = E\left [ \Ent(\eoAct(F)) \right ] .
  \ee
  \item \textbf{Entropy reduction} form of mutual information:
  \be
    \Inf(\Act;F) = \Ent(\Act) - \Ent(\Act|F).
  \ee
  \item \textbf{Data processing inequality}:
  For any transformation $g(F)$ of a random variable or vector $F$,
  \be
    \Inf(\Act;g(F)) \leq \Inf(\Act;F).
  \ee
  \item \textbf{Chain rule} of mutual information:
  \be
    \Inf(\Act; (F,G)) = \Inf(\Act; F) + \Inf(\Act;G | F).
  \ee
\end{itemize}

\setcounter{equation}{0}
\section{Proofs}
\label{sec:proofs}


For ease of reference, we begin by restating our notation and assumptions.
\bals
  \out_\period, \para, \Act_\period &\in \mathbb{R}^\options &\textrm{Outcome, parameter, and action vectors}\\
  A_t \in \feasActs &\subseteq \{\act\in \{0,1\}^\options:\, \|\act\|_1 = \batch\} &\textrm{Feasible allocations and batch size}\\
  \out_{\option \period} &\in [0,1] & \textrm{Bounded outcomes}\\
  \para &=\Exp_\period[\out_\period|\para] & \textrm{Parameters are expectation of outcomes}\\
  \epara_\period &=\Exp_\period[\para] = \Exp_\period[\out_\period] &\textrm{Prior expectation of the parameter (at $\period$)}\\
  \R(\act) &=\Exp_\period[\langle \act, \out_\period \rangle | \para]= \langle \act, \para \rangle &\textrm{Linear (combinatorial) expected rewards}\\
  \out_\period(\act) &=  (\act_{\option}\cdot \out_{\option \period} :\; \option = 1,\ldots,\options) & \textrm{Observable outcomes (semi bandit)}\\
  \oAct &\in \argmax_{\act \in \feasActs} \R(\act) = \argmax_{\act \in \feasActs} \langle \act, \para \rangle & \textrm{Optimal action}\\
  \eparaopt_{\option \period} &= \Exp_\period[\para_\option | \oAct_\option=1] = \Exp_\period[\out_{\option \period} | \oAct_\option=1] & \textrm{Conditional expectation of parameters} \\
  \eoAct_\period &= \Exp_\period[\oAct] &\textrm{Expected optimal action}
\eals

For Thompson sampling we have that $\Act_\period$ has the same distribution as $\oAct$, and therefore
$$
  \Exp_\period[\Act_\period] = \Exp_\period[\oAct] = \eoAct_\period.
$$

We next prove three preliminary Lemmas, before combining them in the proof of Theorem \ref{theo:bounding_thompson} itself.

\begin{lem}[Bounding regret by the component-wise information]
  \label{lem:bounding_componentwise}
  $$\Exp_\period[\R(\oAct) - \R(\Act_\period)]\leq \sqrt{\frac{\options}{2} \cdot \sum_{\option=1}^\options \eoAct_{\option \period}^2 \cdot \KL \left (\eparaopt_{\option \period}, \epara_{\option \period}  \right ) }.$$
\end{lem}

\textbf{Proof of Lemma \ref{lem:bounding_componentwise}:}
\bal
\setcounter{proofeq}{\value{equation}}
  \Exp_\period[\R(\oAct) - \R(\Act_\period)] &= \Exp_\period[\langle \oAct - \Act_\period, \para \rangle]\\
  &= \langle \eoAct_\period , \eparaopt \rangle - \langle \eoAct_\period , \epara_\period \rangle\\
  &\leq \sqrt{\options \cdot \sum_{\option=1}^\options \eoAct_{\option \period}^2 \cdot \left (\eparaopt_{\option \period} - \epara_{\option \period}  \right )^2 }\\
  &\leq \sqrt{\frac{\options}{2} \cdot \sum_{\option=1}^\options \eoAct_{\option \period}^2 \cdot \KL \left (\eparaopt_{\option \period}, \epara_{\option \period}  \right ) }
\eal
These steps hold for the following reasons.
\begin{enumerate}[(B1)]
  \setcounter{enumi}{\theproofeq}
  \item By definition of $\R$.
  \item By splitting the inner product, and using (i) iterated expectations, conditioning on $\oAct_\option=1$ for each component $\option$ in turn,
  and (ii) independence of $\Act_\period$ and $\para_\period$ and the definition of Thompson sampling.
  \item By Cauchy Schwarz (for the inner product with a $\options$-vector of $1$s).
  \item By Pinsker's inequality
  , applied to Bernoulli random variables with expectation $\eparaopt_{\option \period}, \epara_{\option \period}$.
\end{enumerate}
$\Box$\\[10pt]


\begin{lem}[Divergence and component-wise information gain]
  \label{lem:boundingKL}
  $$\eoAct_{\option \period}^2 \cdot \KL \left (\eparaopt_{\option \period}, \epara_{\option \period}  \right )
  \leq \Inf_\period(\oAct_\option; \out_\period(\Act_\period), \Act_\period).$$
\end{lem}

\textbf{Proof of Lemma \ref{lem:boundingKL}:}\\
For the purpose of this proof, construct a Bernoulli random variable $\widetilde{\out}_{\option \period}$ with expectation $\out_{\option \period}$, independently of everything else.
Note that $\Exp_\period[\widetilde{\out}_{\option \period}] = \epara_{\option \period}$.
$ \KL \left (\eparaopt_{\option \period}, \epara_{\option \period}  \right )$ can be interpreted as the KL-divergence between 
the distribution of $\widetilde{\out}_{\option \period}$ conditional on $\oAct_\option = 1$ and
the (unconditional) distribution of $\widetilde{\out}_{\option \period}$.
Taking the expectation over $\oAct_\option$ of the KL-divergence yields the mutual information between $\oAct_\option$ and $\widetilde{\out}_{\option \period}$,
$\Inf_\period(\oAct_\option; \widetilde{\out}_{\option \period})$:
\bal
  \Inf_\period(\oAct_\option; \widetilde{\out}_{\option \period}) =
  \eoAct_{\option \period} \cdot &\KL \left (\Exp_\period[\para_{\option \period} | \oAct_\option=1], \epara_{\option \period}  \right ) \nonumber \\
 +   (1-\eoAct_{\option \period}) \cdot &\KL \left ( \Exp_\period[\para_{\option \period} | \oAct_\option=0], \epara_{\option \period}  \right ),
  \label{eq:component_info}
\eal
and thus
\bal
\setcounter{proofeq}{\value{equation}}
  \eoAct_{\option \period}^2 \cdot \KL \left (\eparaopt_{\option \period}; \epara_{\option \period}  \right )
  &\leq \eoAct_{\option \period} \cdot \Inf_\period(\oAct_\option; \widetilde{\out}_{\option \period})\\
  &\leq \eoAct_{\option \period} \cdot \Inf_\period(\oAct_\option; {\out}_{\option \period})\\
  &= \Inf_\period(\oAct_\option; \Act_{\option \period} \cdot \out_{\option \period}, \Act_{\option \period})\\
  &\leq \Inf_\period(\oAct_\option; \out_\period(\Act_\period), \Act_\period).
\eal
These steps hold for the following reasons.
\begin{enumerate}[(B1)]
  \setcounter{enumi}{\theproofeq}
  \item Because the second term in Equation \eqref{eq:component_info} is non-negative.
  \item By the data-processing inequality, applied to the mapping from ${\out}_{\option \period}$ to $\widetilde{\out}_{\option \period}$.
  \item By the law of iterated expectations, applied to $\Inf_\period(\oAct_\option; \Act_{\option \period} \cdot \out_{\option \period}, \Act_{\option \period})$, averaging over the distribution of $\Act_{\option \period}$ (under Thompson sampling).
  \item By the data processing inequality, again.
\end{enumerate}
$\Box$\\[10pt]

\begin{lem}[Bounding the sum of component-wise information]
  \label{lem:bounding_sum}
  $$\sum_{\period = 1}^\periods \sum_{\option=1}^\options \Inf_\period(\oAct_\option; \out_\period(\Act_\period), \Act_\period)
  \leq  \batch \cdot \left [ \log\left ( \tfrac{\options}{\batch} \right ) +1 \right ] 
  $$
\end{lem}

\textbf{Proof of Lemma \ref{lem:bounding_sum}:}
\bal
\setcounter{proofeq}{\value{equation}}
  \sum_{\period = 1}^\periods \sum_{\option=1}^\options \Inf_\period(\oAct_\option; \out_\period(\Act_\period), \Act_\period)
  &= \sum_{\option=1}^\options \Inf_1(\oAct_\option; (\out_\period(\Act_\period), \Act_\period:\; \period=1,\ldots,\periods))
  \\
  &\leq  \sum_{\option=1}^\options \Ent_1(\oAct_\option)\\
  &= - \sum_{\option=1}^\options \left [\eoAct_{\option,1}  \log(\eoAct_{\option,1}) + (1-\eoAct_{\option,1}) \log(1-\eoAct_{\option,1}) \right ] \\
  &\leq \options \cdot \left (\tfrac{\batch}{\options}\log\left ( \tfrac{\options}{\batch} \right ) +
  \left (\tfrac{\options- \batch}{\options} \right ) \log\left (\tfrac{\options} {\options-\batch}\right )  \right ) \\
  &\leq \batch \cdot \left [ \log\left ( \tfrac{\options}{\batch} \right ) +1 \right ] 
\eal
These steps hold for the following reasons.
\begin{enumerate}[(B1)]
  \setcounter{enumi}{\theproofeq}
  \item The chain rule of mutual information.
  \item The entropy reduction form of mutual information 
  and the non-negativity of (conditional) entropy.
  \item The definition of entropy for $\oAct_\option$.
  \item Jensen's inequality.
  \item The inequality $\log(1+x) \leq x$ for $x=m/(d-m)$.
\end{enumerate}
$\Box$\\[10pt]

\textbf{Proof of Theorem \ref{theo:bounding_thompson}:}~\bal
\setcounter{proofeq}{\value{equation}}
  \Exp_1 \left [ \sum_{\period = 1}^\periods \left ( \R(\oAct) - \R(\Act_\period)  \right ) \right ]
  &=\Exp_1 \left [ \sum_{\period = 1}^\periods \Exp_\period \left [  \R(\oAct) - \R(\Act_\period) \right ]  \right ]\\
  &\leq \Exp_1 \left [ \sum_{\period = 1}^\periods \sqrt{\frac{\options}{2} \sum_{\option=1}^\options \Inf_\period(\oAct_{\option}; \out_\period(\Act_\period), \Act_\period) }  \right ]\\
  &\leq  \sqrt{ \frac{1}{2} \options \periods \Exp_1 \left [ \sum_{\period = 1}^\periods\sum_{\option=1}^\options \Inf_\period(\oAct_{\option}; \out_\period(\Act_\period), \Act_\period)   \right ]}\\
  &\leq \sqrt{\frac{1}{2}\options \periods  
  \batch \cdot \left [ \log\left ( \tfrac{\options}{\batch} \right ) +1 \right ] }.
\eal
These steps hold for the following reasons.

\begin{enumerate}[(B1)]
  \setcounter{enumi}{\theproofeq}
  \item The law of iterated expectations.
  \item Lemma \ref{lem:bounding_componentwise}.
  \item Cauchy-Schwarz for the inner product with a $\periods$-vector of $1$s.
  \item Lemma \ref{lem:bounding_sum}.  
\end{enumerate}
$\Box$\\[10pt]



\end{document}